\documentclass[fleqn,usenatbib]{mnras}

\usepackage{newtxtext,newtxmath}

\usepackage[T1]{fontenc}

\DeclareRobustCommand{\VAN}[3]{#2}
\let\VANthebibliography\thebibliography
\def\thebibliography{\DeclareRobustCommand{\VAN}[3]{##3}\VANthebibliography}

\usepackage{siunitx}

\usepackage{graphicx}	
\graphicspath{{./Figs/}}
\usepackage{amsmath}	
\usepackage{lipsum}

\newcommand*{\CaLine}{Ca\,\textsc{ii} \SI{854.2}{\nano\metre}}
\newcommand*{\Ha}{H$\alpha$}

\newcommand*{\Lw}{\textit{Lightweaver}}

\newcommand*{\AbsDL}{|\Delta\lambda|}
\DeclareSIUnit\erg{erg}

\title[Flare Kernels May be Smaller than You Think]{Flare Kernels May be Smaller than You Think: Modelling the Radiative Response of Chromospheric Plasma Adjacent to a Solar Flare.}

\author[Osborne \& Fletcher]{
Christopher M.\,J.\,Osborne$^{1}$\thanks{E-mail: \href{mailto:christopher.osborne@glasgow.ac.uk}{christopher.osborne@glasgow.ac.uk}},
Lyndsay Fletcher$^{1,2}$
\\
$^{1}$SUPA School of Physics and Astronomy, Kelvin Building, University of Glasgow, G12 8QQ, UK\\
$^{2}$Rosseland Centre for Solar Physics, University of Oslo, P.O. Box 1029 Blindern, NO-0135, Oslo\\
}

\date{Accepted XXX. Received YYY; in original form ZZZ}

\pubyear{2022}

\begin{document}
\label{firstpage}
\pagerange{\pageref{firstpage}--\pageref{lastpage}}
\maketitle

\begin{abstract}
Numerical models of solar flares typically focus on the behaviour of directly-heated flare models, adopting magnetic field-aligned, plane-parallel methodologies. With high spatial- and spectral-resolution ground-based optical observations of flares, it is essential also to understand the response of the plasma surrounding these strongly heated volumes. We investigate the effects of the extreme radiation field produced by a heated column of flare plasma on an adjacent slab of chromospheric plasma, using a two-dimensional radiative transfer model and considering the time-dependent solution to the atomic level populations and electron density throughout this model. The outgoing spectra of \Ha{} and \CaLine{} synthesised from our slab show significant spatial-, time-, and wavelength-dependent variations (both enhancements and reductions) in the line cores, extending on order \SI{1}{\mega\metre} into the non-flaring slab due to the incident transverse radiation field from the flaring boundary. This may lead to significant overestimates of the sizes of directly-heated flare kernels, if line-core observations are used. However, the radiation field alone is insufficient to drive any significant changes in continuum intensity, due to the typical photospheric depths at which they forms, so continuum sources will not have an apparent increase in size. We show that the line formation regions near the flaring boundary can be driven upwards in altitude by over \SI{1}{\mega\metre} despite the primary thermodynamic parameters (other than electron density) being held horizontally uniform. This work shows that in simple models these effects are significant and should be considered further in future flare modelling and interpretation.
\end{abstract}




\section{Introduction}

\begin{figure}
    \centering
    \resizebox{1.0\hsize}{!}{\includegraphics{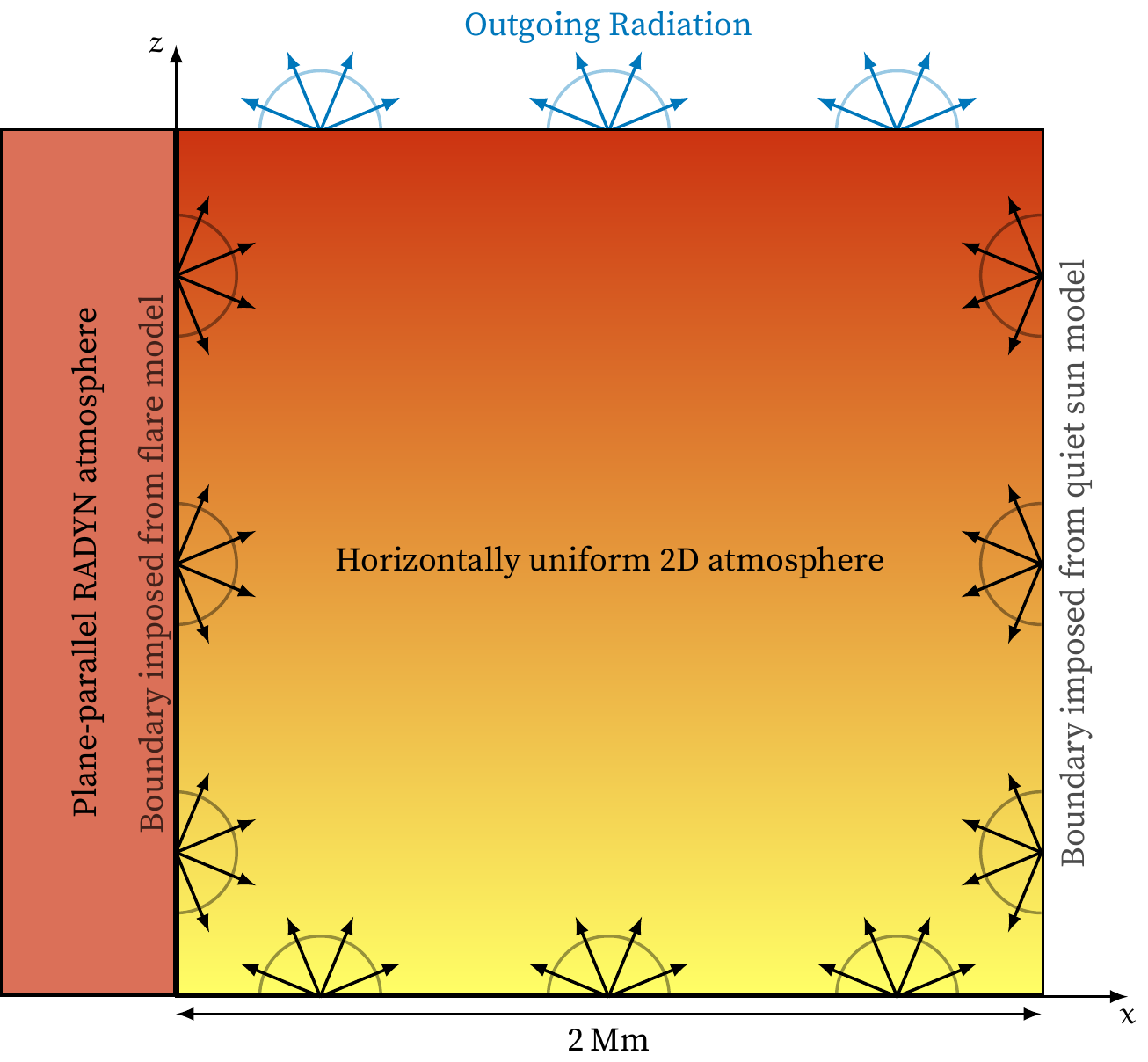}}
    \caption{Configuration of the two-dimensional simulation showing the flaring boundary condition.}
    \label{Fig:SimConfig}
\end{figure}

The modelling of the radiative output of solar flares using radiation hydrodynamic codes such as RADYN \citep{Carlsson1992a,Abbett1999,Allred2015}, HYDRAD \citep{Bradshaw2003,Bradshaw2013,Reep2019}, and FLARIX \citep{Varady2010,Heinzel2015} has become commonplace for the comparison and interpretation of observed chromospheric spectral lines.
These models assume a plane-parallel quasi-one-dimensional magnetic field-aligned view of the flare-heated plasma where, from the perspective of the radiative transfer calculations, each 
plasma column has infinite horizontal extent. 
However, the highest spatially resolved observations of flaring loops suggest that the individual kernels that these radiation hydrodynamic codes seek to model are exceedingly small, on the order of tens to hundreds of \si{\kilo\metre} in diameter \citep{Jing2016}.
Although the plasma surrounding the flare kernel is likely to be significantly cooler due to both the lack of direct heating, and the strongly suppressed cross-field heat conduction relative to that across the field \citep{Spitzer1953},
the photons produced by a flare are not constrained by the magnetic field and will impinge on neighbouring plasma, affecting the conditions therein by radiatively pumping the atomic populations.
This can change the emissivity and opacity of the chromospheric regions neighbouring the flare-heated volumes, as well as localised heating through absorption of radiation. Both will create visible effects on outgoing radiation in an extended region around the directly-heated flare kernel, such that measured flare areas may overestimate the area of the directly-heated kernel. Observed kernel areas are often used in forming estimates of energy fluxes, beam densities etc. It is the purpose of this work to investigate the effect of irradiation of the plasma surrounding a flare kernel on the emergent line intensity from this volume.   

The effects of the flare's radiative flux on nearby plasma has been previously considered with respect to observations of `core-halo' patterns in ground-based line and continuum observations \citep{Neidig1993,Xu2006}, TRACE white light \citep{Hudson2006}, and HINODE/Solar Optical Telescope continuum (G-band) and line (Fe~\textsc{i} \SI{630.2}{\nano\m}) observations \citep{Isobe2007}.
These effects are typically proposed to be due to radiative backwarming \citep[e.g.][]{Metcalf1990a}, in which 
downwards-going radiation emitted isotropically at a higher altitude affects a deeper layer. Limited absorption along the light's path means that it is likely to be able to influence a large area of plasma giving rise to a `halo' backwarmed patch illuminated by a smaller overlying `core'.

In the following we investigate a different geometry, looking at the non-local effects of flaring radiation on an adjacent slab of chromospheric plasma.
Similar problems have been looked at in the quiet Sun.
\citet{Leenaarts2012a} used the \textit{Multi3d} radiative transfer code \citep{Leenaarts2009} to synthesise and investigate the formation of \Ha{} in three-dimensional snapshots of quiet sun atmosphere from the \textit{Bifrost} code \citep{Gudiksen2011}.
They found that a three-dimensional treatment yields significant differences from the plane-parallel column-by-column treatment.
On the other hand, \citet{Bjorgen2019} state that these differences were less significant for the bright structure of a model active region \citep[using the simulation of][]{Cheung2019}, however we comment that the differences are substantial in the quieter regions adjacent to the bright structure when a column-by-column treatment is applied.
These are precisely the regions we are interested in investigating here, and our simple two-dimensional model should highlight whether a more complete multi-dimensional treatment is necessary.

\citet{Stepan2013a} employed the PORTA code \citep{Stepan2013} to investigate the importance of linear scattering polarisation around the border of flare ribbons, using an academic model with a two-level atom.
They found that even with horizontally uniform thermodynamic parameters, inhomogeneities in the radiation field due to changes in non-thermal collisional transition rates at the edge of flare ribbon were capable of producing significant scattering polarisation.

To capture the effects of flaring radiation on adjacent chromosphere we will present, for the first time, fully time-dependent two-dimensional 
models with complete optically thick treatment in which a slab of chromospheric materials responds to incident radiation from an adjacent flare model.
We focus on the outgoing spectra and atomic level populations from this slab whilst leaving a treatment considering the evolution of the temperature in the slab due to absorbed radiation to a future study.
Plane-parallel modelling with the RADYN code by \citet{Carlsson2002}, has highlighted the need to treat the hydrogen populations and plasma ionisation state in a time-dependent manner when considering rapid dynamic changes in the chromosphere due to propagating waves or flares, and we will validate the necessity of this treatment for the case of a radiatively-excited flare-adjacent region.

In Section \ref{Sec:Config}, we will present the setup of our model followed by the spectroscopic variations in the \Ha{} and \CaLine{} spectral lines for two different RADYN flare models in Section \ref{Sec:Results}.
Finally, we analyse the spectral line formation regions to understand the effects of irradiation on the slab and the emergent intensity, as well as demonstrating the need for a complete time-dependent treatment in Section \ref{Sec:Discussion}.

\section{Model Configuration}\label{Sec:Config}

\subsection{Two-Dimensional Slab}

In the following we present models of the radiation emitted by a horizontally uniform slab of quiet Sun chromosphere illuminated by an adjacent RADYN flare model.
The simulation is set up as shown in Fig.~\ref{Fig:SimConfig}: the primary simulation domain is a \SI{2}{\mega\metre} wide slab of plasma set to the parameters of the initial relaxed pre-flare atmosphere used in the associated RADYN simulation.
On the left-hand side of the slab ($x=0$) we place the time-dependent RADYN model, and compute the intensity along each ray of the angular quadrature (discrete ray set) used to integrate the radiation field in the two-dimensional slab, at each depth in the simulation.
The other $x$ boundary is treated equivalently, but remains fixed at the initial quiet Sun atmosphere used in the RADYN simulation and the slab.
The atmosphere is homogenous and infinite along the $y$-axis (perpendicular to the plane of the diagram).
Large domains are computationally costly, so the $\SI{2}{\mega\m}$ width of this domain was chosen by a manual iterative process to ensure that the angle-averaged radiation field in the slab close the quiet Sun boundary did not change too dramatically from its initial value, otherwise this boundary would be spuriously sinking large quantities of energy.

The mass density stratification is fixed in the two-dimensional slab, with the plasma being held static, however we allow the electron density in the slab to vary following charge conservation as this can have a significant impact on the formation of upper chromospheric lines such as \CaLine{}.
Similarly to the original RADYN models, the $z$ stratification of the model changes over time, and is based on a combination of the grids present in the RADYN model for both the initial quiet Sun atmosphere, and the current timestep.
We use 450 points spaced across the entire vertical extent of the RADYN model to ensure that the transition region is well resolved in both the flare model and the adjacent chromospheric slab.
The atomic level populations are interpolated between the $z$ grids used in 
consecutive timesteps 
and are then scaled to match local mass density (to prevent the growth of errors as points move through the transition region).
The 6 rays per octant of the unit sphere angular quadrature of \citet{Stepan2020} was chosen as the plasma in the slab is static, and a 5-ray Gauss-Legendre quadrature is sufficient to resolve the anisotropies in the radiation field in the plane-parallel boundary.
The atomic level populations start from statistical equilibrium, then at each timestep as the radiation from the boundary condition changes, the populations are updated in a time-dependent fashion, and the outgoing intensity is computed.
All of the models shown here are simulated using the \Lw{} framework \citep{Osborne2021}, with boundary conditions treated as coupled radiative transfer models using RADYN's thermodynamics where necessary, thanks to the flexibility of \Lw{}.
In the two-dimensional slab, the BESSER formal solver of \citet{Stepan2013} is used, along with linear interpolation of atmospheric parameters to grid-ray intersections where needed.

The left-most column of the slab immediately adjacent to the RADYN model requires special treatment due to the implementation of the boundary condition: the incoming radiation is fixed for each timestep and right-going ray of the angular quadrature.
As a result, the radiation along these directions is not affected by the local plasma parameters, or directly included in the calculation of the local approximate $\Lambda$ operator.
To mitigate this column appearing unphysically dark compared to both the flare model and the adjacent column to the right, we set its thermodynamic parameters and atomic populations to match those in the plane-parallel flare model, and hold these fixed over the course of each timestep.

The use of the above parameters relating to angular and spatial quadrature is discussed and tested in greater detail in Chapter 6 of \citet{Osborne2021b}.

\subsection{Plane-Parallel Flaring Boundary}

Similarly to the process undertaken in \citet{Osborne2021a}, the plane-parallel flaring boundary is treated in a time-dependent manner, reprocessing the thermodynamic atmospheric properties from the RADYN model, and computing the associated NLTE populations and radiation.
For the models presented here, we do not consider the advection of the atomic populations due to bulk plasma flows in the RADYN model, as its effects on the intensities in these models are relatively small \citep{Kasparova2003}, and instead adopt the same interpolation and scaling approach used in the 2D slab but scaling to the new time-dependent mass density taken from the next timestep of the RADYN model.

The same RADYN model parameters as those used by \citet{Osborne2021a} are used for the flaring boundary conditions.
As discussed there, these models represent ``typical'' RADYN simulations, and lie well inside the parameter space outlined by observations.
We therefore present two models starting from an initial VAL3C atmosphere \citep{Vernazza1981}, heated by an analytic electron beam following the approach of \citet{Emslie1978} with a spectral index $\delta=5$, low-energy cut-off of \SI{20}{\kilo\electronvolt}, and a constant energy deposition for \SI{10}{\s} of either $1\times10^6$ or \SI{1e7}{\J\per\square\m\per\s}.
The models are then allowed to evolve for a further \SI{40}{\s}.
These two models will henceforth be referred to as F9 and F10 respectively, based on their different energy deposition \footnote{In cgs units, as commonly used for RADYN simulations, these correspond to $1\times10^9$ and \SI{1e10}{\erg\per\square\cm\per\s} respectively.}.

\subsection{Model Atoms}

As in \citet{Osborne2021a}, we consider the radiation and level populations of hydrogen and calcium outside of LTE.
For consistency, our model atoms are the same as those used in the F-CHROMA RADYN models\footnote{Produced by the F-CHROMA project and available from
\url{https://star.pst.qub.ac.uk/wiki/doku.php/public/solarmodels/start.}}, consisting of a five-level + continuum model for both species.
The model hydrogen atom has ten bound-bound transitions, and the calcium model has five.
In addition to these models, other atomic species are considered in LTE to provide the background atmospheric opacity and emissivity.
The models for these are taken from the standard RH distribution \citep{Uitenbroek2001}.

\section{Results}\label{Sec:Results}


\begin{figure*}
    \centering
    \resizebox{1.0\hsize}{!}{\includegraphics{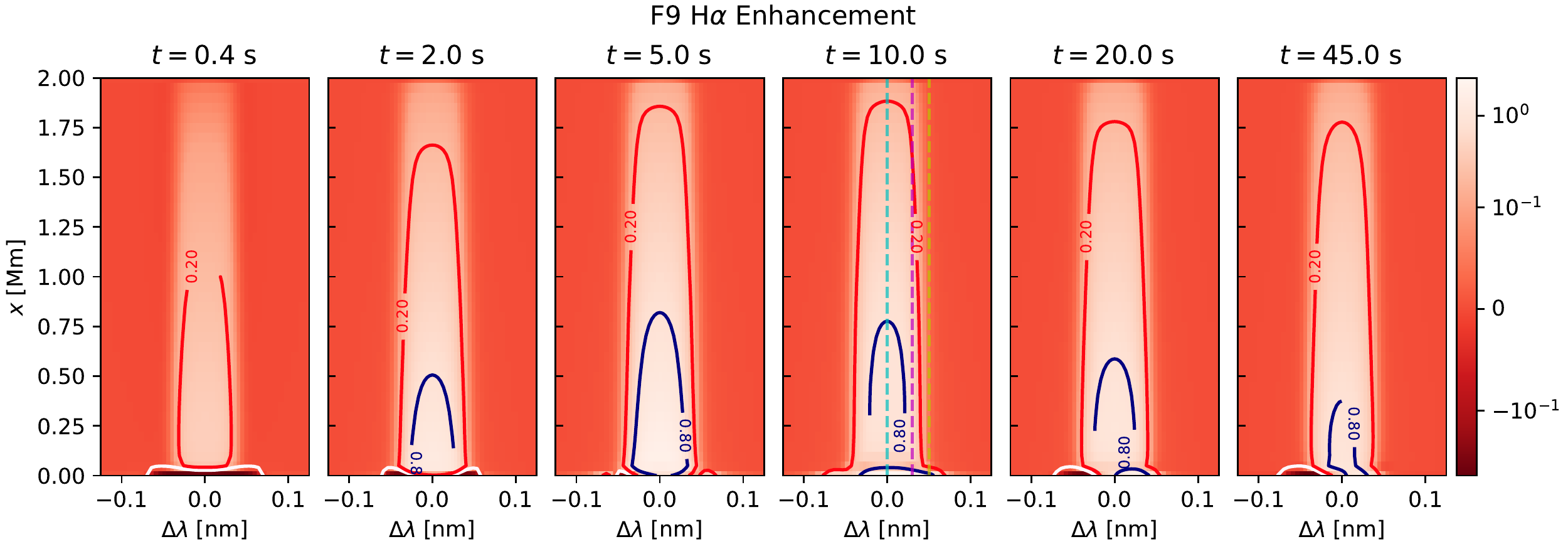}}
    \caption{Relative Enhancement in the H$\alpha$ line profile at different times for the slab illuminated by the F9 model. The three highlighted wavelengths in the $t=\SI{10}{\s}$ panel will be analysed in greater detail throughout the paper.}
    \label{Fig:F9_HalphaMulti}
\end{figure*}

\begin{figure*}
    \centering
    \resizebox{1.0\hsize}{!}{\includegraphics{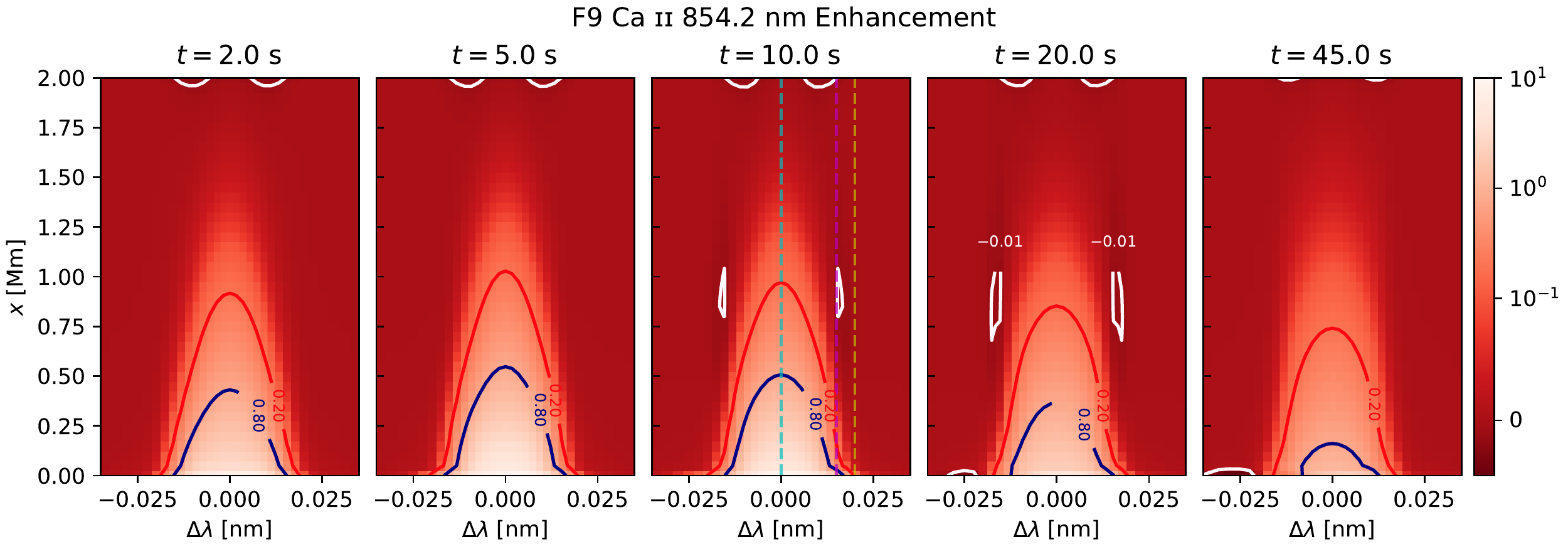}}
    \caption{Relative Enhancement in the \CaLine{} line profile at different times for the slab illuminated by the F9 model.}
    \label{Fig:F9_8542Multi}
\end{figure*}

In Figures \ref{Fig:F9_HalphaMulti} and \ref{Fig:F9_8542Multi} we show the relative enhancement in the outgoing spectrum of \Ha{} and \CaLine{} respectively as a function of distance $x$ from the flaring boundary, synthesised vertically through the atmosphere, at different timesteps in the F9 model.
Overlaid on these figures are contours for:
\begin{itemize}
    \item a 1\% reduction of intensity, in white,
    \item a 20\% enhancement of intensity, in red,
    \item an 80\% enhancement of intensity, in blue.
\end{itemize}
We see that at $t=\SI{10}{\s}$ after flare onset, at the end of beam heating, the \Ha{} line core is enhanced by over 20\% at \SI{1.75}{\mega\metre} from the flaring boundary.
The extent of this 20\% enhancement then remains significant for the remainder of the \SI{50}{\s} simulation.
It is interesting to note that the extent of the 80\% line-core enhancement in \Ha{} reduces between $t=\SI{5}{\s}$ and $t=\SI{10}{\s}$, along with a notable decrease in enhancement around $x=\SI{0.1}{\mega\m}$ despite the continued heating and increase in outgoing radiation from the flaring boundary condition.

The \CaLine{} line is simpler to interpret, showing an increase in enhancement over the course of heating, with a similar aspect at both $t=5$ and \SI{10}{\s}, followed by a slow decay over the remainder of the simulation.
There are small lobes around $\AbsDL{}=\SI{0.02}{\nano\metre}$ at $t=\SI{10}{\s}$ and $t=\SI{20}{\s}$ (shown by white contours) where the line wing intensity drops slightly below its average quiet Sun values.

\begin{figure*}
    \centering
    \resizebox{1.0\hsize}{!}{\includegraphics{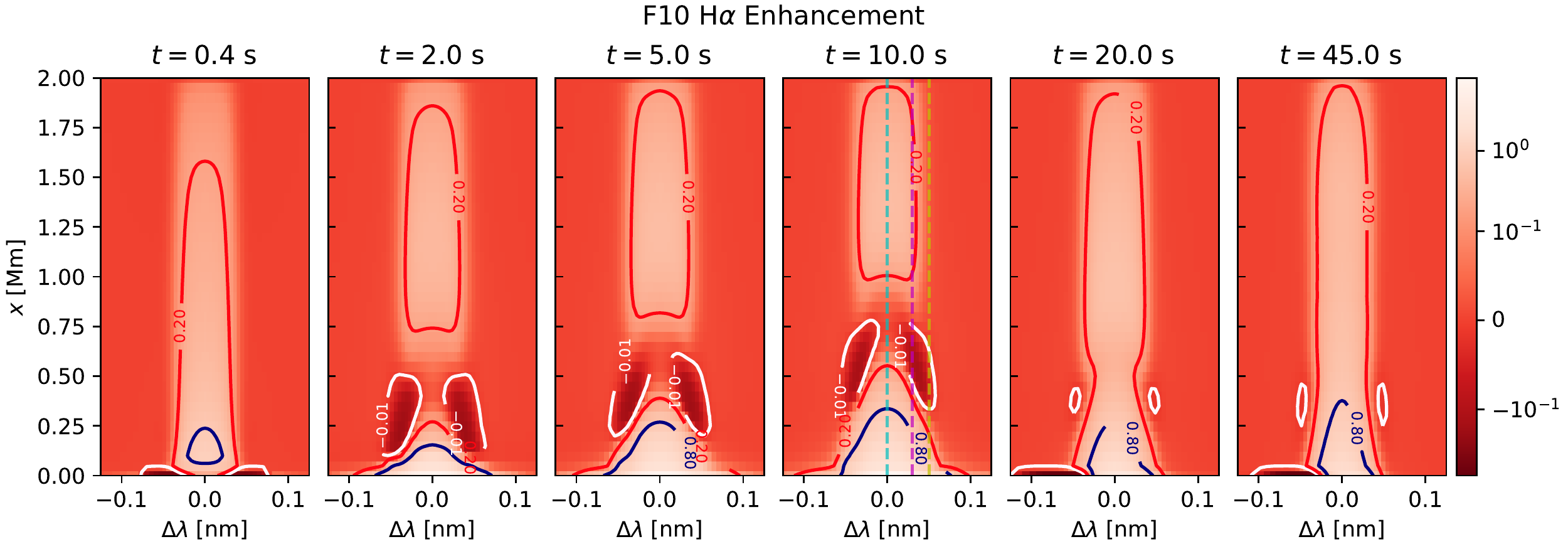}}
    \caption{Relative Enhancement in the \Ha{} line profile at different times for the slab illuminated by the F10 model.}
    \label{Fig:F10_HalphaMulti}
\end{figure*}

\begin{figure*}
    \centering
    \resizebox{1.0\hsize}{!}{\includegraphics{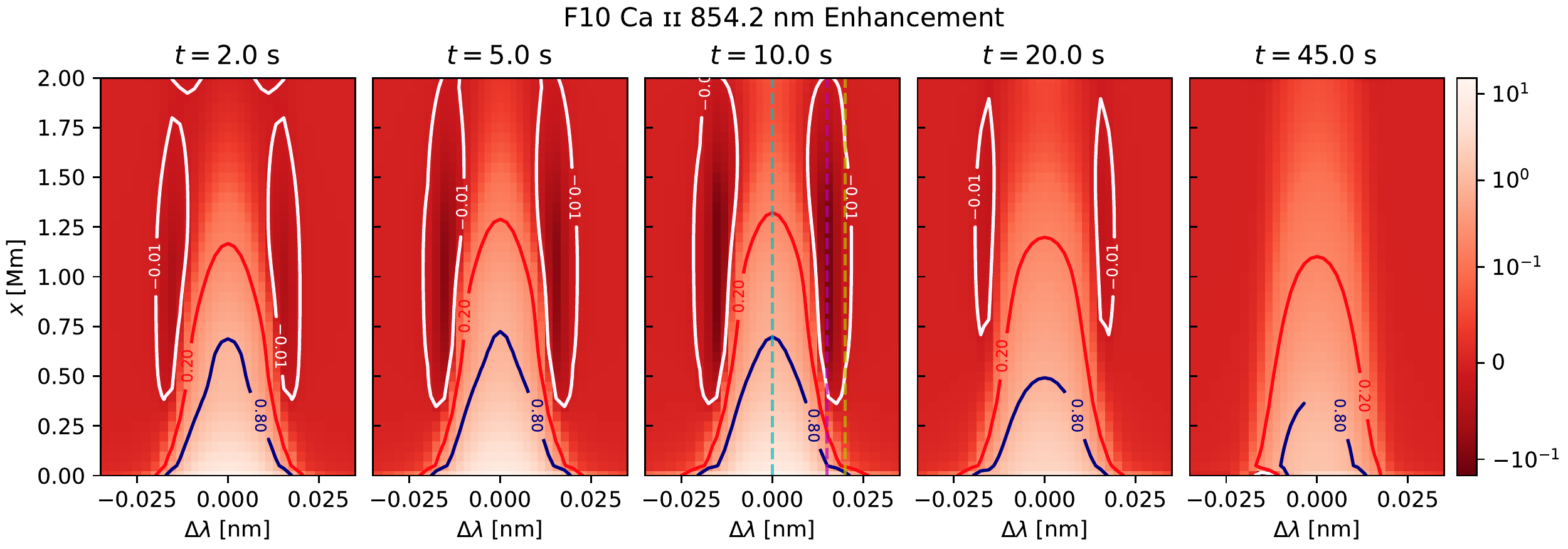}}
    \caption{Relative Enhancement in the \CaLine{} line profile at different times for the slab illuminated by the F10 model.}
    \label{Fig:F10_8542Multi}
\end{figure*}

Figures \ref{Fig:F10_HalphaMulti} and \ref{Fig:F10_8542Multi} show the equivalent enhancements in the \Ha{} and \CaLine{} spectra for the F10 flaring boundary.
There are significant differences in the \Ha{} enhancements in the two models: whilst the 20\% enhancement lobe in the $1-\SI{2}{\mega\metre}$ range remains similar (albeit slightly extended in the F10 model), there are significant differences in the $0-\SI{0.75}{\mega\metre}$ range where significant absorption features are seen in both the red and blue sides of the line core following a strongly enhanced region adjacent to the flaring boundary.
These relative absorption features dissipate almost entirely over the \SI{10}{\s} after the end of the flare heating, and will be analysed in greater detail later.

The \CaLine{} in the F10 model behaves similarly to the F9 case, with more extended enhancements and larger negative enhancements on the edges of the line core.

We note that for both spectral lines, the far wings and continuum ($|\Delta\lambda| \geq \SI{0.1}{\nano\metre}$ for \Ha{} and $|\Delta\lambda| \geq \SI{0.03}{\nano\metre}$ for \CaLine{}) present no notable variation throughout the evolution of these models, and this is true of the continuum intensity in general.

It is important to recognise the scales - 0.5$0-\SI{0.75}{\mega\metre}$ or more -  over which the \Ha{} and \CaLine{} lines are significantly (80\%) enhanced, compared to the source sizes in flares observed at high resolutions. For example, the event studied by \cite{Kuridze2015} has chromospheric sources in the cores of these two lines with structures that are 1--2 arcseconds or less wide, which corresponds to $1.2-\SI{2.4}{\mega\metre}$ when the source position angle is accounted for. These calculations suggest that a significant fraction of the apparent flare kernel in the core could be due to irradiation from the actual flare kernel which is considerably smaller. Interestingly, in this event the sources in the line wings are much more compact, consistent with our finding of no notable variation in the continuum intensity.  


\section{Discussion}\label{Sec:Discussion}

From the previous section, we can see that the scale of the enhancement in the adjacent slab observed in these chromospheric lines is not directly proportional to the beam heating injected into the flare model.
Indeed, the maximum enhancements in both spectral lines are slightly larger for the F10 model than the F9, but the shape of the enhanced regions in the (wavelength, outgoing position) plane are significantly more complex in the former.
The F10 model also presents regions of significant reduction in the outgoing line profiles that in the F9 model are barely present in \CaLine{} and not present at all in \Ha{}.
In the following we will analyse the properties of the slab to investigate how the flare's radiation affects the spectral line forming regions.

\subsection{Spectroscopy and Contribution Functions}

\begin{figure*}
    \centering
    \resizebox{1.0\hsize}{!}{\includegraphics{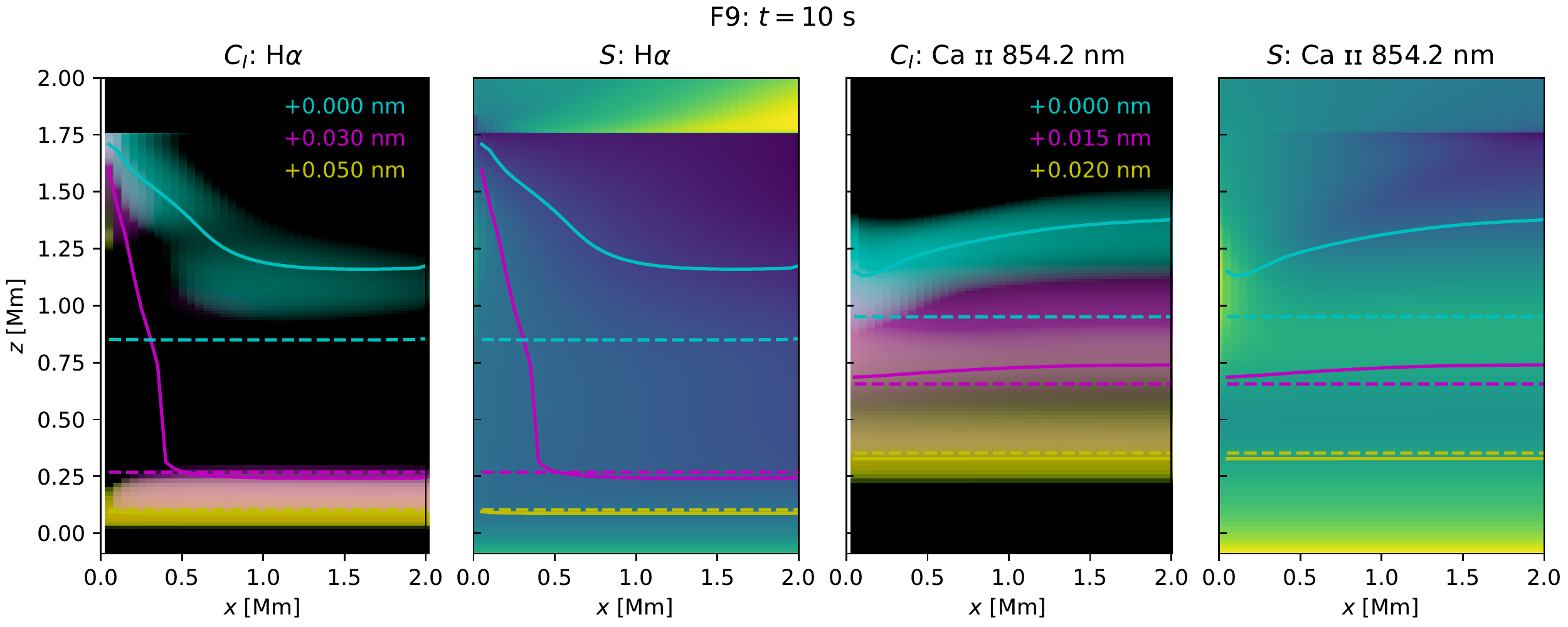}}
    \caption{Contribution function around three different wavelengths and source functions for the \Ha{} and \CaLine{} spectral lines in the slab adjacent to the F9 model at $t=\SI{10}{\s}$. The $\tau_\nu=1$ surface at the associated wavelengths are overplotted as solid lines, and the dashed lines show the $\tau_\nu=1$ layer in the slab at $t=\SI{0}{\s}$.}
    \label{Fig:F9_CO}
\end{figure*}

To understand changes in the formation of spectral lines, it is common to use the contribution function \citep[e.g.][]{Carlsson1997,Kerr2016,Osborne2021a}, the integrand of the formal solution of the radiative transfer equation, to highlight key formation regions.
Given a model with converged NLTE atomic populations, the contribution function can be computed for each line of sight and wavelength.
In our model, due to the vertical synthesis, the final contribution function is computed for each column.
This leads to a 3-dimensional quantity defined over the spatial and wavelength axes of the model.
To understand the important changes that occur in the slab, three wavelengths are highlighted in Figures~\ref{Fig:F9_HalphaMulti}--\ref{Fig:F10_8542Multi}, and we will investigate changes in spectral line formation in the neighbourhood of these wavelengths.

In the first and third panels of Figure~\ref{Fig:F9_CO} we show for the F9 model at $t=\SI{10}{\s}$, in three different colours, the local contribution to the contribution function weighted by a narrow Gaussian around the indicated offsets from the line rest wavelengths.
This is similar to, and inspired by the COCOPLOT technique of \citet{Druett2021}\footnote{We have rotated the colours from red, green, and blue to cyan, magenta, and yellow to aid in colour blind legibility.
As the addition of two of these colours at full brightness can exceed the $[0, 1]$ brightness range representable in standard dynamic range content, we apply a tonemapping step to assist in visualisation.
This tonemapping step, commonly applied to computer-rendered graphics, is based on the Academy Color Encoding System (ACES, \url{https://github.com/ampas/aces-dev}) tonemapper, using a modification of the implementation provided by S. Hill, M. Pettineo, and D. Neubelt on GitHub (\url{https://github.com/TheRealMJP/BakingLab/blob/master/BakingLab/ACES.hlsl})}.
In these panels, the solid cyan, magenta and yellow lines indicate the $\tau_\nu=1$ surface, at the associated wavelength at $t=\SI{10}{\s}$ and the dashed lines indicate this surface at $t=\SI{0}{\s}$ in the initial atmosphere.
The apparent discontinuity around $z=\SI{1.75}{\mega\m}$ is due to the transition region and the coronal region above this is not important for the formation of these chromospheric lines.
For \Ha{}, the line core contribution function (cyan) next to the flaring boundary ($x=\SI{0}{\mega\metre}$), moves from an altitude of $z=\SI{0.9}{\mega\metre}$ in the initial atmosphere to $z=\SI{1.7}{\mega\metre}$.
The displacement of the line formation region for the inner wing ($\AbsDL{}=\SI{0.03}{\nano\metre}$) is even more extreme, moving from $z=\SI{0.3}{\mega\metre}$ to $z=\SI{1.6}{\mega\metre}$.
There is a less significant effect further into the line wing ($\AbsDL{}=\SI{0.05}{\nano\metre}$), with a contribution visible immediately adjacent to the flare at an altitude of $z=\SI{1.3}{\mega\metre}$, but the $\tau_\nu=1$ surface for this wavelength does not move appreciably from the quiet Sun case.
The $\tau_\nu=1$ levels for $\AbsDL{}=0,\,\SI{0.03}{\nano\metre}$ fall off rapidly with distance from the flaring boundary, with the latter returning to close to the quiet Sun case around $x=\SI{0.5}{\mega\metre}$.
The former of these settles to $z=\SI{1.2}{\mega\metre}$ around $x=\SI{1}{\mega\metre}$ from the flaring boundary, that is, the line core continues to form approximately \SI{0.3}{\mega\metre} higher than in the quiet Sun case, for the entire $x=[0,1]\,\si{\mega\metre}$ range.

We note that it is, of course, inappropriate to discuss line-formation height as a single value per column of the atmosphere, as there is an expansive and diffuse region that contributes to its formation: instead, we are describing the evolution of the altitude of the $\tau_\nu=1$ layer, which serves as a proxy ``mean'' line formation height.
The displacement of  the $\tau_\nu=1$ surfaces is not as significant for \CaLine{}, but it is interesting to note that it increases \emph{with} distance from the flaring boundary for both the line core and $\AbsDL{}=\SI{0.015}{\nano\metre}$.
The other two panels in this figure show the line source function for these two lines, with these same $\tau_\nu=1$ lines overplotted.
We see that at a constant depth, the source function is enhanced over its quiet sun values close to the flaring boundaries between an altitude of $z=\SI{0.75}{\mega\metre}$ and $z=\SI{1.6}{\mega\metre}$.

\begin{figure*}
    \centering
    \resizebox{1.0\hsize}{!}{\includegraphics{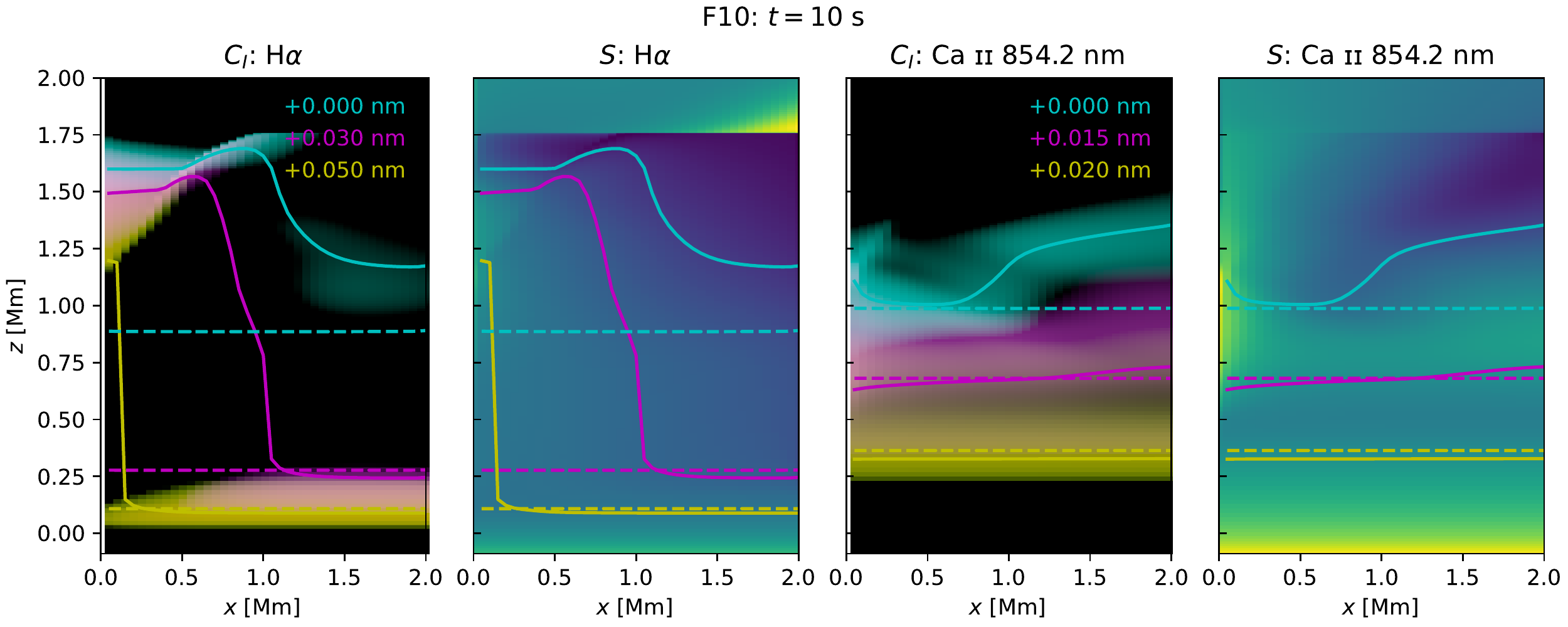}}
    \caption{Contribution functions and source functions in the slab illuminated by the F10 flare model (equivalent to Figure \ref{Fig:F9_CO}).}
    \label{Fig:F10_CO}
\end{figure*}

Figure~\ref{Fig:F10_CO}, similarly to Figure~\ref{Fig:F9_CO}, shows the contribution functions and source functions of the \Ha{} and \CaLine{} spectral lines in the F10 model at $t=\SI{10}{\s}$.
Looking first at the \Ha{} contribution functions, we see that the enhanced contribution the the line core and $\AbsDL{}=\SI{0.03}{\nano\metre}$ from the upper chromosphere is enhanced much further into the slab than in the F9 case.
The line core $\tau_\nu=1$ layer has moved from $z=\SI{0.9}{\mega\metre}$ to $z=\SI{1.6}{\mega\metre}$, and remains at this altitude for the first \SI{0.5}{\mega\metre} of the slab.
Over the following $0.5-\SI{1}{\mega\metre}$ range the $\tau_\nu=1$ line rises smoothly to $z=\SI{1.7}{\mega\metre}$.
This line then smoothly drops down to an altitude of $z=\SI{1.2}{\mega\metre}$ where it remains until the far boundary, similarly to the F9 case.

The $\AbsDL{}=\SI{0.03}{\nano\metre}$ $\tau_\nu=1$ surface behaves somewhat similarly, rising to an altitude of $z=\SI{1.5}{\mega\metre}$ for the first $\SI{0.5}{\mega\metre}$ of the model, before rising slightly then dropping down to slightly below its original formation altitude over the following \SI{0.5}{\mega\metre} range.
The altitude of the $\tau_\nu=1$ surface for $\AbsDL{}=\SI{0.05}{\nano\metre}$ is significantly increased for the first \SI{100}{\kilo\metre} of the model before falling back to the formation height in the original model.


\subsection{Eddington-Barbier Interpretation}

To better interpret the source function and the formation of the outgoing line profiles, we can adopt the Eddington-Barbier approximation.
This approximation is derived for a source function varying linearly with optical depth in a semi-infinite atmosphere, however it often proves relatively accurate in more general NLTE line-formation situations \citep[e.g.][]{Leenaarts2012a}.
The Eddington-Barbier approximation states that $I_\nu(\tau_\nu=0, \mu) \approx S_\nu(\tau_\nu=\mu)$, where the left-hand term is the outgoing intensity, and $S_\nu$ is the source function.
In this work we are only considering vertical rays (i.e. $\mu=1$), so by this approximation, it is the variation of the source function at $\tau_\nu=1$ that should describe the variations in outgoing intensity.
In the following we will qualitatively compare the outgoing line variations to those predicted by the Eddington-Barbier approximation.

\begin{figure}
    \centering
    \resizebox{1.0\hsize}{!}{\includegraphics{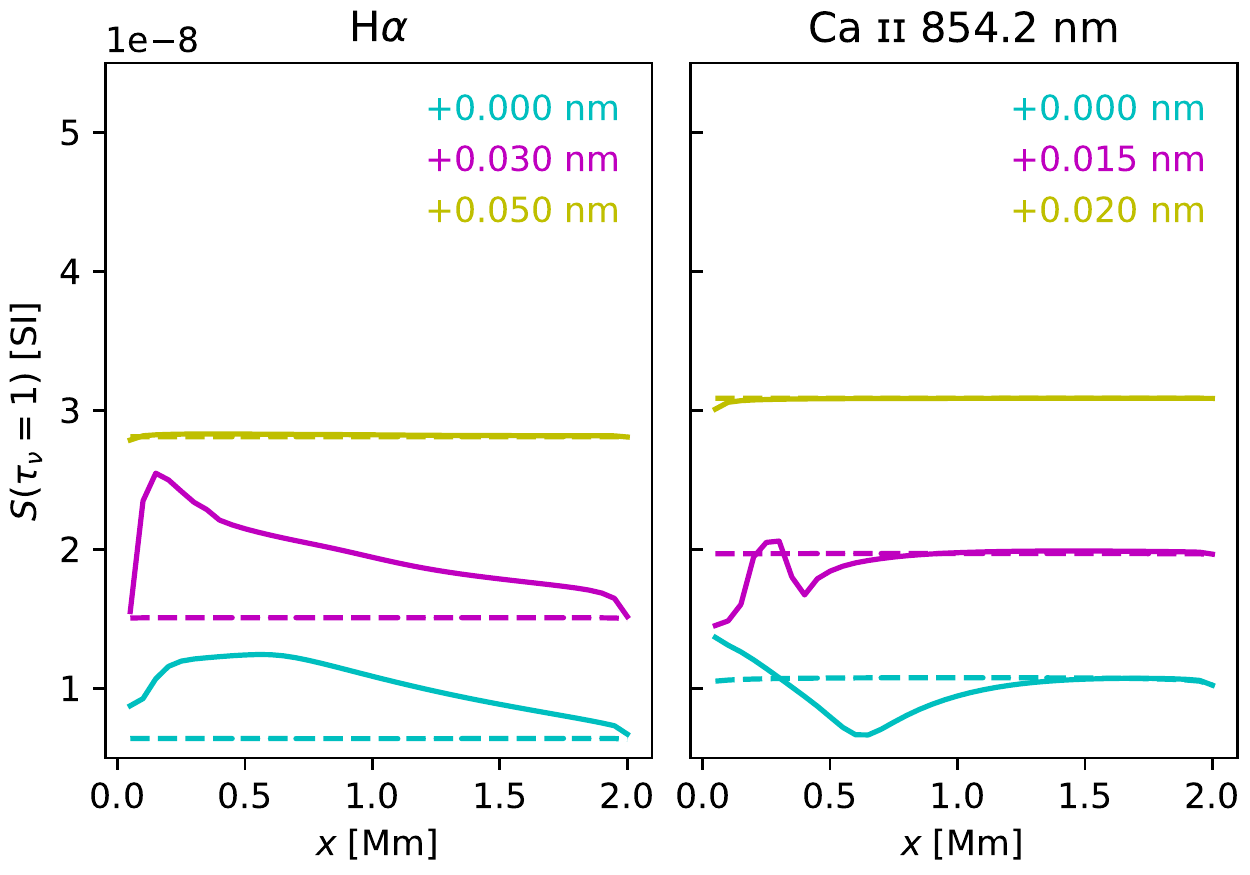}}
    \caption{Source function (\si{\watt\per\square\metre\per\steradian\per\hertz}) values along the $\tau_\nu=1$ line for the different wavelengths shown in the contribution function plots of Figure \ref{Fig:F9_CO} (F9 model). The solid line shows the values at $t=\SI{10}{\s}$, and the dashed line the values along the $\tau_\nu=1$ lines at $t=\SI{0}{\s}$.}
    \label{Fig:F9_Sfn}
\end{figure}

\begin{figure}
    \centering
    \resizebox{1.0\hsize}{!}{\includegraphics{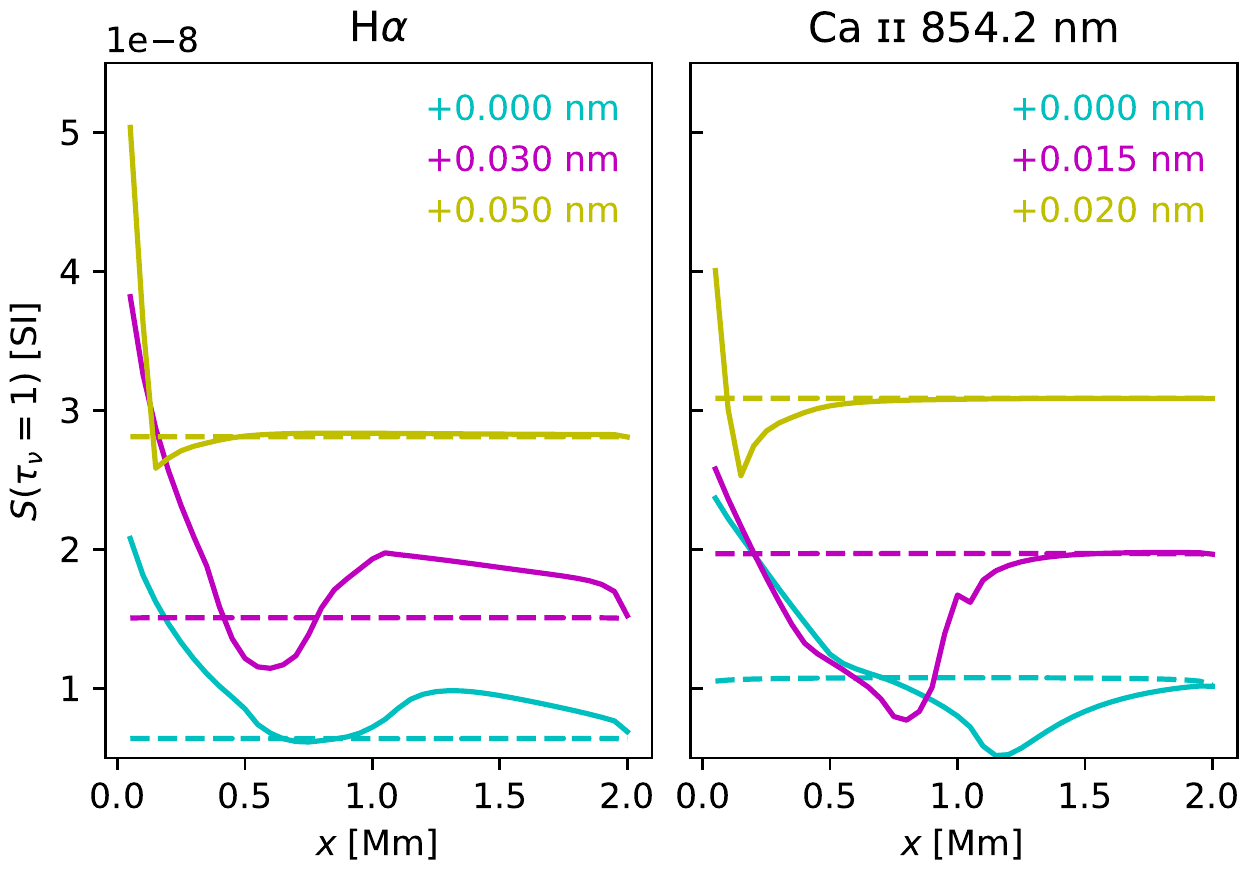}}
    \caption{Source functions (\si{\watt\per\square\metre\per\steradian\per\hertz}) along the $\tau_\nu=1$ lines for the F10 model (equivalent to Figure \ref{Fig:F9_Sfn}).}
    \label{Fig:F10_Sfn}
\end{figure}

In Figures \ref{Fig:F9_Sfn} and \ref{Fig:F10_Sfn}, we show the value of the source function, for the spectral lines and wavelengths considered, along the $\tau_\nu=1$ line depicted in Figures \ref{Fig:F9_CO} and \ref{Fig:F10_CO}, for the F9 and F10 case respectively, both at $t=\SI{10}{\s}$.
The dashed lines show the source function along the $\tau_\nu=1$ lines in the initial slab.
The difference between the solid and dashed lines can then be qualitatively compared with the evolution of the line enhancements shown along the dashed cuts of the $t=\SI{10}{\s}$ panels in Figs.~\ref{Fig:F9_HalphaMulti}--\ref{Fig:F10_8542Multi}.

Looking first at \Ha{} in Figure \ref{Fig:F9_Sfn}, we see that the source function of the line increases as $\AbsDL{}$ increases.
This is to be expected from an absorption line, and indicates some basic plausibility of this approach.
As discussed in relation to Figure \ref{Fig:F9_CO}, the source function at $\AbsDL{}=\SI{0.05}{\nano\m}$ does not change appreciably, however there are strong enhancements close to the flaring boundary for the line core and $\AbsDL{}=\SI{0.03}{\nano\m}$ source functions.
The peak in the $\AbsDL{}=\SI{0.03}{\nano\m}$ source function is sharper and occurs closer to the flaring boundary than for the line core.
This can be plausibly compared to the spectroscopic data in Figure \ref{Fig:F9_HalphaMulti}, as the enhancement in the line core is significantly more extended in $x$, whereas the enhancement at $\AbsDL{}=\SI{0.03}{\nano\m}$ rises to a similar value, but drops off much more rapidly.

The \CaLine{} source function slices in Figure \ref{Fig:F9_Sfn} do not align so well to the spectroscopic data.
The line core starts enhanced, before dropping below its starting value to a minimum around \SI{0.6}{\mega\metre}.
There is also some oscillatory behaviour in the $\AbsDL{}=\SI{0.015}{\nano\metre}$ source function.
Neither of these correspond well to the behaviour of the spectral line, which is enhanced close to the flaring boundary, and then decreases with distance from this boundary.
A possible explanation for this is the extent of the line forming region: as shown in Figure \ref{Fig:F9_CO} the regions with significant contribution in each wavelength band have a notably larger extent in $z$ for \CaLine{} than they do for \Ha{}.
Due to the larger regions contributing to the emission of \CaLine{} at these wavelengths, it is reasonable that the Eddington-Barbier approximation performs less well, especially when comparing with the variation of the source function, which is far from linear with optical depth.

In Figure \ref{Fig:F10_Sfn} we show the equivalent variation of the source function in the F10 model.
For the \Ha{} line, the line core source function is enhanced close to the flaring boundary, then falls to close to its original value between $x=0.5$ and \SI{1}{\mega\metre}, before becoming slightly enhanced over the remaining region.
The source function at $\AbsDL{}=\SI{0.03}{\nano\metre}$, behaves similarly, but drops below its original value between $x=0.5$ and \SI{0.8}{\mega\metre}.
After the excursion of the $\tau_\nu=1$ line to $z=\SI{1.2}{\mega\m}$ close to the flare, where the source function is strongly enhanced, the source function of the $\AbsDL{}=\SI{0.05}{\nano\m}$ line remains below its initial value until $x=\SI{0.5}{\mega\m}$.
Once again, the Eddington-Barbier approximation well explains the outgoing spectroscopic view of the line profile shown in Figure \ref{Fig:F10_HalphaMulti}: for the line core, the intensity is enhanced over the first \SI{0.5}{\mega\m} of the model, then falls back to no net enhancement over the $x=0.5-\SI{0.85}{\mega\m}$ range, and is then visibly enhanced over the rest of the simulation.
The intensity at $\AbsDL{}=\SI{0.03}{\nano\m}$ behaves similarly, but with a significant reduction between $x=0.5-\SI{0.75}{\mega\m}$.
Finally, the intensity along the $\AbsDL{}=\SI{0.05}{\nano\metre}$ line is enhanced close to the flare, is notably reduced between $x=0.3-\SI{0.6}{\mega\m}$, and remains close to its original value for the rest of the model.

For \CaLine{}, the Eddington-Barbier approximation correctly predicts enhancements over the first \SI{0.5}{\mega\metre} of the model, however for the line core, it predicts a reduction for the remainder of the simulation, which is not observed.
At $\AbsDL{}=\SI{0.015}{\nano\metre}$, it predicts a reduction between $x=0.2$ and \SI{1.2}{\mega\m}, whereas this reduction is primarily observed between $x=0.5$ and \SI{1.5}{\mega\metre}.

Thus, for both of these flare models we find that the Eddington-Barbier approximation appears to accurately describe the behaviour of \Ha{} in the slab, but performs poorly on \CaLine{}.
This is likely due to the more complex variation of the source function and the larger contributing regions in each column of the \CaLine{} line.


\subsection{Electron Density}

\begin{figure}
    \centering
    \resizebox{1.0\hsize}{!}{\includegraphics{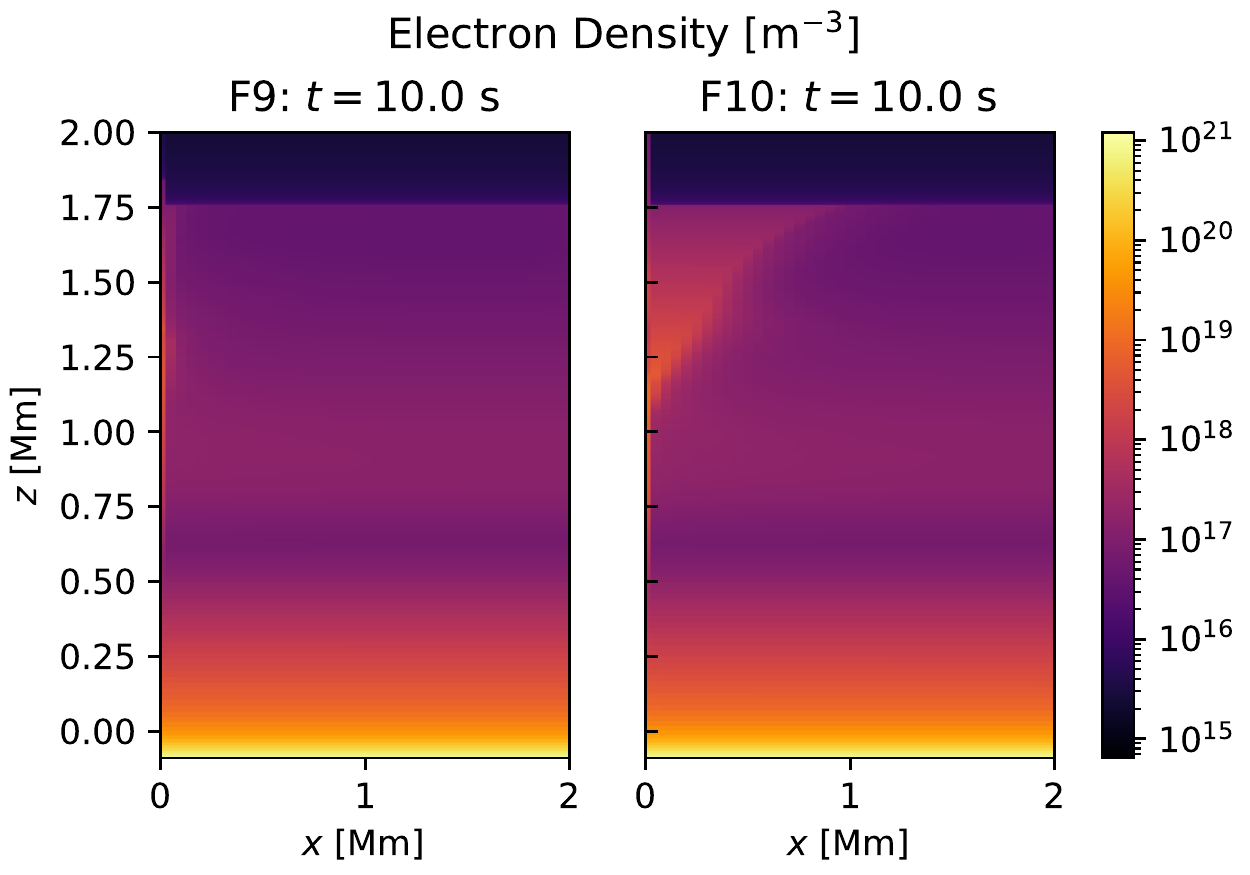}}
    \caption{Electron density in the slab adjacent to both flare models at $t=\SI{10}{\s}$.}
    \label{Fig:Nes}
\end{figure}

In Figure \ref{Fig:Nes} we show the electron density in the slab at $t=\SI{10}{\s}$ for both the F9 and F10 models.
For the F9 model, the only noticeable variation is present around $z=\SI{1.25}{\mega\m}$ close to the flaring boundary. Throughout the rest of the model the electron density is close to its value at the quiet sun boundary.
In the F10 model, there is an approximately triangular region of enhancement between $z=\SI{1.2}{\mega\m}$ at the flaring boundary and $z=\SI{1.75}{\mega\m}$, $x=\SI{1}{\mega\m}$.
As the thermodynamic parameters of the slab are held constant, we can see that the increased radiation from the more energetic flare model is capable of creating significant photoionisation up to \SI{1}{\mega\m} from the flaring boundary.

\subsection{Statistical Equilibrium Comparison}

\begin{figure}
    \centering
    \resizebox{1.0\hsize}{!}{\includegraphics{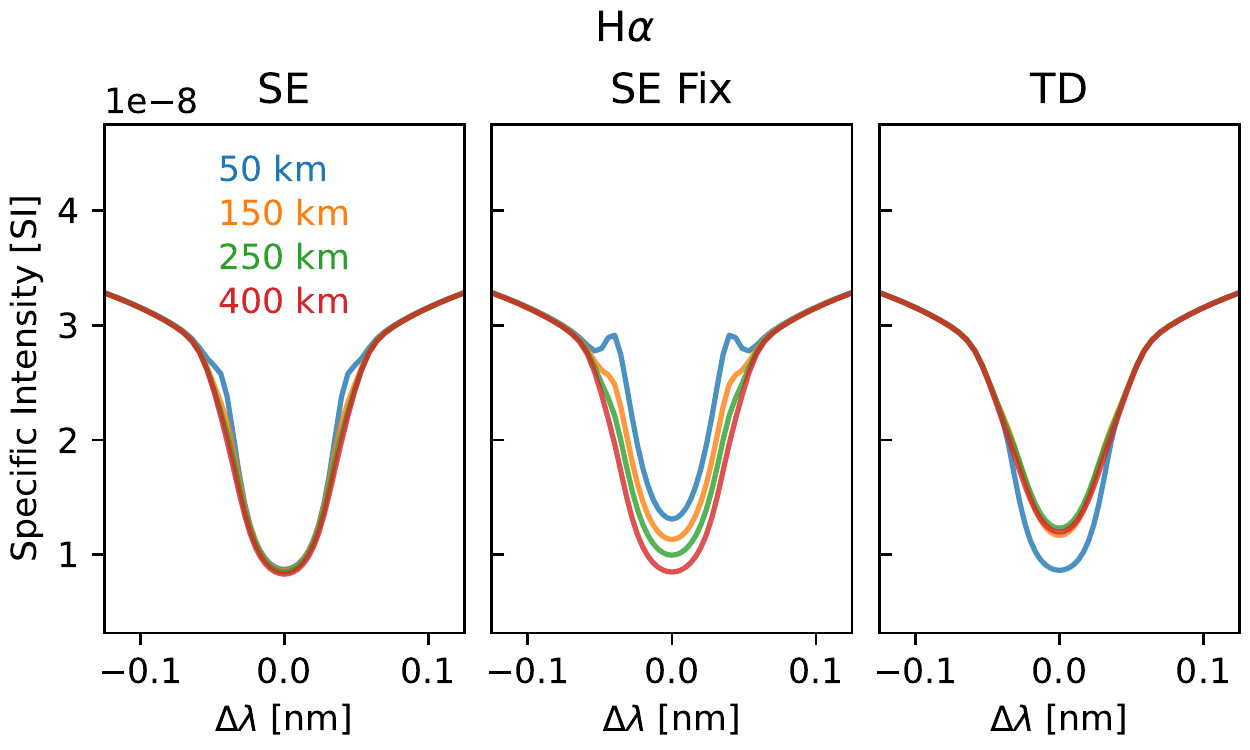}}
    \caption{Comparison of emergent \Ha{} line profiles (\si{\watt\per\square\metre\per\steradian\per\hertz}) from the F9 model at $t=\SI{15}{\s}$ at different slices in $x$ under the treatments of statistical equilibrium with charge conservation (SE, left-most panel), statistical equilibrium with electron density loaded from the time-dependent model (SE Fix, central panel), and the full time-dependent solution (TD, right-hand panel).}
    \label{Fig:HaSeComp}
\end{figure}

\begin{figure}
    \centering
    \resizebox{1.0\hsize}{!}{\includegraphics{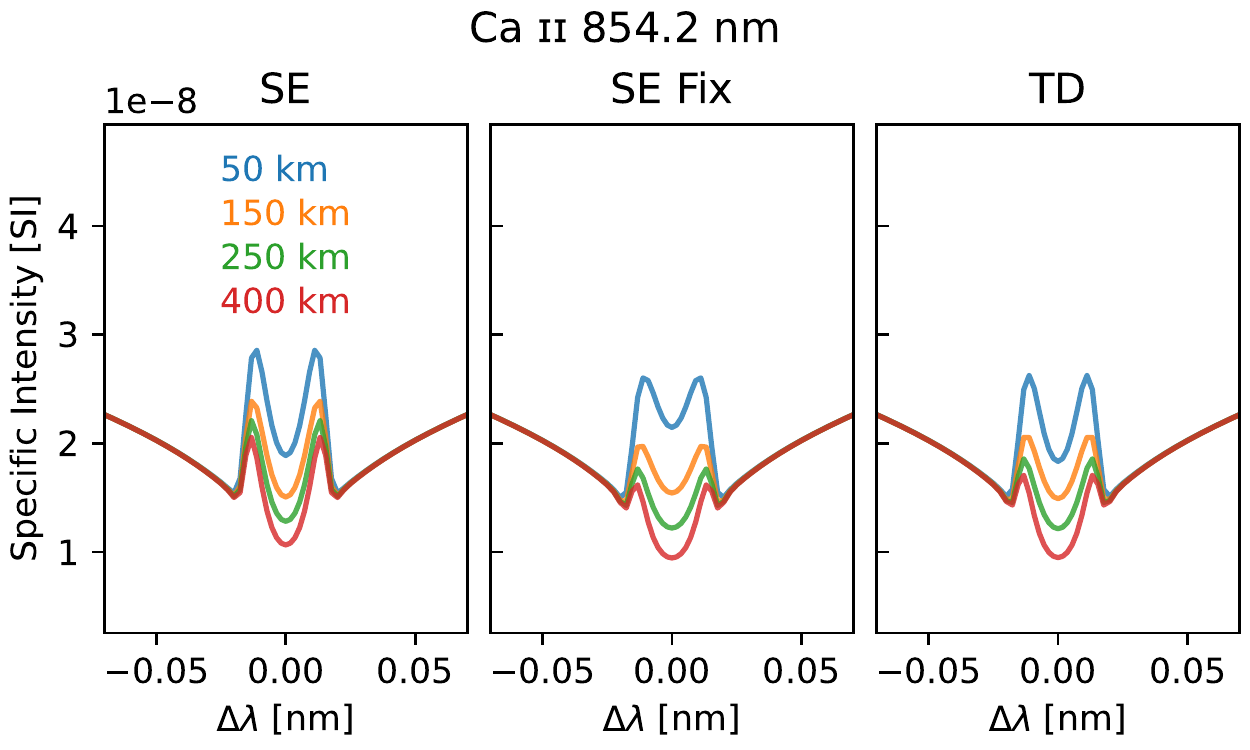}}
    \caption{Comparison of emergent \CaLine{} line profiles (\si{\watt\per\square\metre\per\steradian\per\hertz}) at different slices in $x$ under statistical equilibrium, statistical equilibrium with fixed electron density, and time-dependent models (equivalent to Figure \ref{Fig:CaSeComp}).}
    \label{Fig:CaSeComp}
\end{figure}

In the previous sections we have presented and analysed for the first time the time-dependent radiative output of a two-dimensional slab adjacent to a flare model.
It is commonly accepted that plane-parallel models of the chromosphere must be treated in a time-dependent manner, outside of statistical equilibrium \citep[e.g.][]{Carlsson2002}, due to the very long times needed for the hydrogen ionisation and level populations needed to settle (on the order of 1000s of seconds), however it is not immediately clear whether this will apply to plasma perturbed purely by a transverse radiation field.
In Figures \ref{Fig:HaSeComp} and \ref{Fig:CaSeComp} we show a comparison of the \Ha{} and \CaLine{} spectral lines synthesised at different cuts in $x$ (between $x=50$ and $\SI{400}{\kilo\metre}$) from the F9 model at $t=\SI{15}{\s}$ under three different assumptions.
From left to right we show the synthesis in statistical equilibrium with charge conservation (SE), statistical equilibrium with the electron density taken from the time-dependent model (SE Fix), and the full time-dependent output (TD).
This timestep was chosen as there were difficulties obtaining convergence for the statistical equilibrium model with fixed electron density for $t=\SI{5}{\s}$ and $t=\SI{10}{\s}$.
For \Ha{}, shown in Figure \ref{Fig:HaSeComp}, there are dramatic differences between the three treatments.
The pure statistical equilibrium treatment does not capture the spatially varying enhancement in the line core, whereas the statistical equilibrium with fixed electron density significantly overestimates this enhancement, and those around $\AbsDL{}=\SI{0.05}{\nano\m}$.

For \CaLine{} the differences between the three treatments are less dramatic, but there are still significant differences in line shape between the (left-hand) statistical equilibrium treatment and the other two that are likely due to the change in electron density between the models.
The shape of \CaLine{} is very sensitive to the electron density in the line core formation region \citep[e.g. see the example in][]{Osborne2021,Bjorgen2019}, and it is likely that this is the primary origin of the differences, as the statistical equilibrium with fixed electron density (central panel), agrees well with the full time-dependent results, with slight discrepancies in line core depth over the first \SI{100}{\kilo\m} of $x$.
This agrees well with the results found in time-dependent plane-parallel models presented in Section 5.5 of \citet{Osborne2021b}.

It is clear that the full time-dependent treatment is needed to capture the correct evolution of \Ha{} and the electron density needed to evaluate \CaLine{}.

\section{Conclusion}

We have presented novel simulations of the time-dependent radiation output of a two-dimensional slab of plasma with fixed quiet Sun temperature structure irradiated by an adjacent flaring model.
This represents a ``zeroth-order model'' of the real situation and is likely an upper limit of the effect on the adjacent chromosphere (which will not be truly quiet in practice), but allows for the investigation of the importance of these radiative effects.
We find that this irradiation can create significant enhancements in the \Ha{} and \CaLine{} line profiles over \SI{1}{\mega\m} from the flaring boundary, due to induced changes in the atomic level populations and electron density. 
Given that observed flare sources in these lines can be only a couple of \SI{}{\mega\m} across these enhancements could lead to a significant overestimate of the area of the directly heated flare kernel.
These changes occur in the mid and upper chromosphere, and as a result no changes in continuum intensity are observed in the slab, leading continuum sources to appear comparatively compact, which is consistent with observed behaviour in line core and wing observations.
We also found that the peak enhancement from the slab in these lines did not relate simply to the beam flux injected into the flaring boundary despite the other parameters remaining fixed.
As a result traditional plane-parallel inversion techniques applied to this region are likely to be led to incorrect conclusions regarding the thermodynamic structure of the slab, predicting a significantly hotter region due to the effects of the transverse radiation field, whilst this atmosphere is instead the modified VAL C7 model of \citet{Vernazza1981} used as RADYN's initial atmosphere with the electron density allowed to vary to maintain charge conservation.

We also find that it is necessary to treat the hydrogen populations in a fully time-dependent manner, so as to correctly capture the ionisation balance in the slab and the specie's slow return to statistical equilibrium after a radiative perturbation.
This suggests that to improve correctness, radiative models of active regions must treat hydrogen in a time-dependent manner, vastly increasing the numerical cost of synthesis as each timestep of the model must be iterated in turn.

We also comment that these effects vary as a function of wavelength, with enhancements in the line-core typically extending hundreds of kilometres further into the slab than those in the wings.
This could have implications on the calculation and usage of filling factors as the ``spatial-smearing'' of intensity is not uniform across wavelength, however further modelling will be needed to begin to determine any form of general relation.

As the deluge of high spatial- and spectral-resolution data from ground based telescopes continues, it becomes increasingly important to characterise the effects of the transverse radiation field to allow for the correct interpretation of observations, and truly unlock the meaning of data contained within.
Future models should therefore consider additional physics by both placing the flare model within the two-dimensional slab, possibly using a model with cylindrical symmetry to treat the flare as a cylindrical flux tube, and allowing the chromospheric plasma to evolve through flows and conduction in response to the radiative energy input.
Embedding the flare within the two-dimensional slab will also allow the investigation of the effects of spectral line formation within a compact region of heated atmosphere, rather than the infinite horizontal extent implicitly assumed in plane-parallel models.

\section*{Acknowledgements}

CMJO acknowledges support from the University of Glasgow College of Science and Engineering and the UK Research and Innovation's Science and Technology Facilities Council (STFC) doctoral training grant ST/R504750/1.
LF acknowledges support from the UK Research and Innovation's STFC under grant award number ST/T000422/1.
This research arose in part from discussions held at and around a meeting of the International Space Science Institute (ISSI) team: “Interrogating Field-Aligned Solar Flare Models: Comparing, Contrasting and Improving” organised by G.S. Kerr and V. Polito and we would like to thank ISSI-Bern for supporting this team.
The authors are grateful to the reviewer for suggesting improvements to the manuscript.

\section{Data Availability}

The data underlying this article is available in Zenodo at \url{https://zenodo.org/record/6382792}, and the source code is available on GitHub at \url{https://github.com/Goobley/MsLightweaver2d} (v1.0.0) with archival on Zenodo at \url{https://zenodo.org/record/5484184}.
The plotting scripts (including tonemapping steps) and intermediate data products are available at \url{https://zenodo.org/record/7041606}.


\bibliographystyle{mnras}
\bibliography{Refs} 






\bsp	
\label{lastpage}
\end{document}